# The role of information asymmetry in the market for university-industry research collaboration[1]

*Giovanni Abramo[1], Ciriaco Andrea D'Angelo[2], Flavia Di Costa[2] and Marco Solazzi[2]*

**1 Corresponding author.** Affiliation**:** Italian National Research Council and Laboratory for Studies of Research and Technology Transfer - School of Engineering, Dept of Management - University of Rome "Tor Vergata" – Via del Politecnico 1, 00133 Roma – Italy
abramo@disp.uniroma2.it
tel./fax +39 06 72597362

2 Affiliation**:** Laboratory for Studies of Research and Technology Transfer - School of Engineering, Dept of Management - University of Rome "Tor Vergata" – Via del Politecnico 1, 00133 Roma – Italy

# The role of information asymmetry in the market for university-industry research collaboration

**Abstract**

This study concerns the market for research collaboration between industry and universities. It presents an analysis of the population of all Italian university-industry collaborations that resulted in at least one international scientific publication between 2001 and 2003. Using spatial and bibliometric analysis relating to scientific output of university researchers, the study shows the importance of geographic proximity in companies' choices of university partner. The analysis also reveals inefficiency in the market: in a large proportion of cases private companies could have chosen more qualified research partners in universities located closer to the place of business.

## 1. Introduction

The capacity to develop new knowledge is one of the most significant factors in a nation's economic growth, together with the ability to transfer knowledge to the national industrial system. European Union countries, in spite of excellent scientific results, are noted for their lack of success in translating knowledge into useful market innovations (EC 1995). The phenomenon is especially evident in nations such as Italy, where a very large share of scientific research is conducted in the public sector (47.5%); where the industrial system concentrates on low to mid-level technology and industries consist mostly of micro, small and mid-sized companies.

Public research can have a significant and pervasive effect in the sphere of private enterprise (Cohen et al. 2002), and collaboration in research is one of the more important forms of knowledge transfer. Many nations enact policies and stimuli to support such collaboration and transfer, with the European Community Research Framework Programmes being one of the most notable initiatives.

Public–private research collaboration can be considered as an exchange relationship in which both parties obtain benefits (Meyer-Krahmer and Schmoch 1998). For university researchers, collaboration with private business provides additional financing for research and/or complementary physical assets. Private enterprises use collaboration to solve specific technical or design problems, develop new products and processes, conduct research leading to new patents, recruit university graduates and access cutting-edge research (Lee 2000).

In many countries, when a private company needs a university research partner, making the optimal choice is difficult because of the lack of information concerning the scientific excellence of potential partners. Thus, choices are often based on factors related to geographic or social proximity[2]. The present study is intended to cast light on some of the determinants of the choice of research partner and investigate whether market inefficiency occurs in public-private research collaboration. In an efficient market the choice by a private company would be based on the quality of product offered, meaning the scientific knowledge in the pertinent discipline, and its cost. The cost of research in partnership should be directly correlated to the geographic distance between demand and supply, all else being equal. Furthermore, if adequate information concerning the variable of quality,is lacking it is logical that businesses will resort to decisions based on social proximity. Geographic and social proximity thus should play a determining role in the choice of research partner on the part of the company.

The current paper explores the role of geographic proximity among the determinants of the company's choice of university research partner. One striking peculiarity of the Italian context is the information asymmetry on the research quality of potential academic partners; this study aims at verifying whether this causes inefficiency in the market of university-industry research collaboration. The field of observation consists of all the research collaborations in the hard sciences between universities and companies located in Italy that have resulted in at least one scientific publication in the period 2001 to 2003. As will be shown, the bibliometric approach we take enables observation of a large population of public-private collaborations in Italy, certainly more than a patent analysis, which would also be difficult to realize, or an empirical

---

[2] The term "social proximity" is used in the sense given by Boschma (2005): "Social proximity is defined here in terms of embedded relations between agents at the micro-level. Relations between actors are socially embedded when they involve trust based on friendship, kinship and experience".

study based on surveys.

The next section of this paper presents a brief analysis of the literature on the proximity effect in public-private research collaboration. Section 3 describes the basic assumptions and stages involved in the research, while section 4 shows the methodology and the dataset used. Section 5 reports general data describing public-private collaborations. Section 6 presents the results of the analysis of the geographic proximity effect. Section 7 explores the inefficiency in the market for research collaboration due to information asymmetry concerning the quality of university partners. The concluding section comments on the results of the study and indicates possible directions in future research.

**2. The proximity effect in public-private research collaboration**

The concept of geographic proximity has been analyzed in many studies, and has proven so rich and versatile as to enter into a number of policy concepts for regional development. Many national and regional development programs draw on concepts such as proximity between universities and high-tech companies, proximity of researchers or of related economic sectors (Lang 2005).

The concept of geographic limits to the diffusion of knowledge can help explain different levels of growth and development between regions. We see that the presence of excellent universities has a favorable impact on innovation in the surrounding region (Jaffe 1989; Jaffe et al. 1993), while empirical research demonstrates that knowledge flows from the public to private sector diminish with geographic distance (Arundel and Geuna 2004). In general, the number of collaborations between pairs of partners declines exponentially with increasing distance (Katz 1994).

Boschma (2005) studies the relationship between proximity and innovation, evaluating geographic and other dimensions, including cognitive, organizational, social and institutional proximity. Geographic proximity can compensate for the absence of other forms of proximity, thus permitting collaborations between organizations of diverse backgrounds, missions and organizational structures, as is the case between universities and private companies. Geographic proximity should increase the possibility of personal contacts, useful for transfer of tacit knowledge, but it is not alone sufficient for development of collaborations. Other types of proximity can favour collaboration between organizations separated by great distances.

The importance of geographic proximity is undergoing change as transport costs and information and communications technology also continually change (Cairncross 1997; Antonelli 2000). There is a debate on the impact of new technologies, including to what extent they could displace the geographic proximity effect by facilitating scientific collaboration amongst groups spread over wide distances (Handy 1995; Townsend et al. 1998; Hallowell 1999; Roberts 2000; Olson and Olson 2000; Shapiro and Narian 2002).

There has also been some study of the manner in which decision-makers' considerations of the scientific excellence and geographic location of potential partners may be related. When a company that needs a university partner does not have information on excellence the decision-makers will often resort to social proximity, meaning that they select from scientists tied to their personal circle of acquaintances (Prabhu 1999). Access to university resources is partially based on personal contacts of an informal nature, through conferences, workshops, etc., and on the personal contacts of recent graduates that have newly arrived in private companies (Lindelof and Lofsten 2004; Meyer-Krahmer and Schmoch 1998).

Lee and Mansfield (1996) conducted a survey of relations between a sample of 66 high-tech American firms and universities, which again showed that geographic distance results as an important factor in determining collaborations. Other factors such as university excellence did enter into play, but companies tended to finance research in universities that could be found within a radius of 100 miles, even if these did not demonstrate high levels of excellence. Universities situated beyond this range had a greater possibility of being chosen only if they were among those with an optimal reputation. In such cases, the projects involved were usually "basic" research, in which proximity is less determining and scientific prestige of the faculty assumes greater importance.

The study presented in this paper explores the interplay between scientific excellence and geographic distance in the selection of university research partners, in the Italian context.

## 3. Assumptions and stages of this study

The study begins from an assumption that in the absence of information asymmetry concerning the scientific excellence of universities, a private company would behave as a rational decision-maker and choose research partners in relation to their research quality and the costs, in terms of geographic and social distance..

The study is articulated in three stages. First we will identify the pertinent characteristics of the collaborations being examined. In a second step we demonstrate the occurrence of proximity effect in companies' selection of a university partner, by comparing average geographic distances seen in actual collaborations with the distances that could be expected. The expected distance is measured by weighting geographic distances by additional factors that could also contribute to selection of a university collaboration partner: number of scientists, scientific excellence of the university in general and in the specific discipline in which the collaboration occurs. In other words, we assume a market with complete information. The results confirm a marked geographical proximity effect, i.e. the average geographic distance of research partners is shorter than the expected one.

The proximity effect seen could be due to two reasons. The first is that Italian companies, in a cost-benefits analysis, give greater weight to the costs variable and so to geographic distance. The second is that there is information asymmetry: a shortage of information on the quality of research carried out by universities, research groups and single scientists. Before proceeding to a description of the third stage of our study, a review of the Italian context can explain the background for the hypothesis of information asymmetry. The general framework in Italy is an academic system consisting entirely of public universities. The first national research evaluation for the university system was not carried out until 2006 (i.e. after the period under observation). Government funding (approximately 90% of university budgets) is not allocated on the basis of merit. Another feature is that professional recruitment takes place only through a process of nationally administered examination of potential recruits which does not permit any direct competition between universities for candidates. Academics' salaries are not based on merit, rather on rank and seniority. Overall, competition between universities is practically inexistent, whether resources are from government or from other sources (with the exception of a very low fraction of financing coming from national and international agencies that manage large research programs). In such a context, companies lack a basis for perception or classification of objective merit and related data concerning universities and their research groups. It can therefore be expected that such perceptions become highly uncertain. Furthermore, in Italian industry there is little diffusion of bibliometric databases on research output, which could provide more transparent and accessible understanding of public research and new knowledge on offer..

The third stage of this study (starting with Section 7) provides an analysis to determine whether companies' selection of closer research partners is due to information asymmetry or to the weight they place on costs. If companies give weight to costs greater than to quality we would expect to see that there is no higher quality university or research group closer to the company than the one actually chosen. In Section 7 we consider further information about research performance of universities and single scientists which permits us to assess whether a company could have chosen a better university or a better scientist within a shorter distance. In doing so, we are able to prove that information asymmetry is a cause of inefficiency in the market for university-industry research collaboration.

## 4. Methodological approach and dataset

To investigate the phenomenon of public-private research collaboration the study considers scientific articles co-authored by universities and private companies and published in international journals. An analysis of patent applications filed by both private companies and universities could provide a useful complement but in Italy the number of patents filed by all universities is notoriously low. Using the Espacenet search engine we have identified 284 patents filed by Italian universities between 2001 and 2003[3]. Of these, only 20 were co-filed with private companies. Because co-filing does not necessarily imply co-authorship, the number of university patents stemming from research collaboration with industry may be even less than 20. Many university researchers who file patents would also wish to publish scientific articles on the subject, therefore the analysis of publications could capture information concerning patents filed. For these reasons we have chosen not to include patents in our analysis.

The bibliometric approach presents certain limits. The most noted are that scientific collaborations do not always lead to publication of results and that co-authorship of a publication does not necessarily indicate real collaboration (Melin and Persson 1996, Katz and Martin 1997, Laudel 2001). However, publication in co-authorship remains one of the most tangible and best documented indicators of research collaboration (Subramanyam 1983; Katz and Martin 1997, Glänzel and Schubert 2004). Subramanyam (1983) argues that using co-authorship as a proxy for collaboration offers a series of advantages because it is: i) invariant, ii) easily and inexpensively ascertainable iii) quantifiable and iv) non-reactive (i.e. the process of ascertaining collaboration does not affect the process of collaboration itself). According to Glänzel and Schubert (2004), "…almost every aspect of scientific collaboration networks can be reliably tracked by analyzing co-authorship networks by bibliometric methods". Given the paucity of university patents in Italy, co-authorship of scientific publications results as the most representative indicator for the study of research collaboration. Bibliometric analysis offers the advantage of a much broader field of observation compared not only to patent analysis, but also compared to other empirical methods based on partial surveys.

---

[3] After the introduction of academic privilege in 2001, a higher number of patents filed by academics themselves are certainly there, but relevant data are not readily available, which makes their identification really difficult.

The field of observation for this work consists of all 78 Italian universities and all private companies located in Italy. The source for data is the Observatory of Public Research (ORP), managed by the authors, which is based on information from Thomson Reuters Web of Scinece. The information for the current study is based on the $SCI^{TM}$ for the years 2001 to 2003.

The authors used the $SCI^{TM}$ to identify publications (articles and reviews) authored by organizations based in Italy. From this group, the next step was to select those articles co-authored by universities and companies. To do this, the authors identified and reconciled the different ways in which the same organization was indicated in the $SCI^{TM}$ "address" field[4].

Finally, through a "disambiguation" algorithm, each publication was accurately attributed to its respective university author. The Italian university system imposes that every research scientist adhere to a particular scientific disciplinary sector (SDS), therefore it was also possible also to link each publication to the authors' SDSs. The Italian system further groups the SDSs into disciplinary macro-areas[5]. The field of observation for the present analysis includes 8 of the 9 "hard science" disciplines: Mathematics and computer sciences, Physics, Chemistry, Earth sciences, Biology, Medicine, Agricultural and veterinary sciences, Industrial and information engineering[6]. These disciplines include 183 SDSs. The dataset is composed of the $SCI^{TM}$-listed publications for 2001 to 2003 in these 183 SDSs, executed in co-authorship by universities and private companies located in Italy. This includes a total of 1,534 such publications. Out of a total of 78 universities, 10 of them had no research scientist in the SDSs under observation. Of the remaining 68, 58 were found to produce articles in co-authorship with industry in the scientific disciplines under observation. The number of private companies collaborating was 483.

To permit analysis of the potential presence of proximity effect on collaboration, the authors identified the geographic coordinates of each university and company in the dataset. Some collaborations included in the census involved multiple locations of a single company: these were considered as distinct entities. For this reason there are a total of 509 company locations[7] in the dataset.

### 4.1 Working definition of collaboration

---

[4] For details see Abramo et al., 2008a.
[5] For details see http://www.miur.it/atti/2000/alladm001004_01.
[6] The discipline of Civil engineering and architecture was not included because the SCI™ does not provide full coverage of research output.
[7] Unless otherwise specified, from this point on the term "company" includes the multiple locations of the same legal entity.

In this section we describe two ways of analysing research collaborations between universities and industry. These are related to the two distinct levels of analysis in this study: the organizational level (university-company) and the more detailed level of the single disciplinary sector (SDS-company).

The first analysis is at the level of university-company collaboration. By university-company collaboration we mean a research collaboration between a university and a private company (both located in Italy) that has resulted in exactly one co-authored publication of the dataset under consideration. A generic publication by *m* universities and *n* private companies therefore corresponds to *m\*n* university-company collaborations. The 1,534 co-authored publications observed involved 1,983 university-company collaborations. These same 1,534 publications were produced by 1,226 university-company pairs[8].

The analysis at the university-company level is deepened by further analysis at the SDS-company level. One of the main reasons for a company to collaborate with a university is the opportunity to access the specialized knowledge and competencies of university researchers. The disciplinary sectors of the university researchers can be considered as "macro-competencies", or sets of common competencies for researchers[9]. The authors have conducted a specific analysis in order to investigate the selection of collaboration partners at the sectorial level. For the generic publication *i* of the dataset, one has

$$(\text{Number of collaborations at SDS-company level})_i = \left[ \sum_{k=1}^{m_i} \text{SDS\_coauthors}_k \right] * n_i$$

where:
- $m_i$ = number of universities with at least a researcher co-author of publication *i*;
- $\text{SDS\_coauthors}_k$ = number of SDSs, of university *k*, with at least a researcher co-author of publication *i*;
- $n_i$ = number of private companies co-authors of publication *i*.

As an example of this working definition, an article co-authored by three private companies and five academic scientists from a single university but from two different SDSs would result in 6 SDS-company collaborations. We would also see 6 SDS-company collaborations for the case of an article authored by three private companies and 2 university scientists in the same SDS but affiliated to different universities.

This second analysis shows the occurrence of 2,363 SDS-company collaborations, corresponding to 1,755 distinct SDS-company pairs.

In synthesis, in the dataset considered, each publication represents one or more university-company collaborations (collaborations at an organization level) and one or more SDS-company collaborations (collaborations at a sectorial level). Table 1 sums up the calculation of collaborations relating to 1,534 dataset publications.

---

[8] The number of university-company pairs is less than the number of university-company collaborations because, in the period considered, some pairs produced more than one collaboration (publication): certain pairs collaborated more than once. The same consideration will apply to the SDS-company level analysis as well.

[9] In reality, each SDS includes multiple and probably non-quantifiable specializations that evolve in a dynamic manner. However the authors retain that the approximation introduced by considering the SDS as a unique set of competencies, possessed by the participating researchers, permits the understanding of certain aspects of university-industry collaboration.

| Level of analysis | Number of collaborations | Number of pairs |
|---|---|---|
| *University-company* | 1,983 | 1,226 |
| *SDS-company* | 2,363 | 1,755 |

*Table 1: University-company and SDS-company collaborations and pairs in the dataset*

## 5. Characteristics of collaborations

### 5.1 Frequencies of partners and disciplinary sectors per collaboration

Analysis of co-authorships permits identification of the number of partners involved in each collaboration. Table 2 reports the number of publications produced in relation to the number of organizations involved, both in absolute terms and as percentages.

| Number of companies | Number of universities | | | | | |
|---|---|---|---|---|---|---|
| | 1 | 2 | 3 | 4 | 5 | 6 |
| 1 | 1,195 (77.9%) | 228 (14.9%) | 43 (2.8%) | 10 (0.7%) | 3 (0.2%) | 1 (0.1%) |
| 2 | 40 (2.6%) | 7 (0.5%) | 1 (0.1%) | 2 (0.1%) | - | - |
| 3 | 4 (0.3%) | - | - | - | - | - |

*Table 2: Number of publications in relation to numbers of companies and universities in a co-author group; percentage of total publications indicated in parentheses*

The data show that co-authored publications never involve more than 3 companies or 6 universities. The majority of co-authored publications (approximately 78%) involve only two organizations. It is rare (3.5%) that more than one company is involved in a co-author group. However the presence of more than one university in joint research with a private company is more frequent, occurring in 19% of cases. This difference is not surprising, given the nature, different backgrounds and interests of the two types of partners involved.

Table 3 shows the average number of disciplinary sectors seen in collaborations involving different numbers of companies and universities. On average, there are 1.35 SDS involved in each publication..

| Number of companies | Number of universities | | | | | |
|---|---|---|---|---|---|---|
| | 1 | 2 | 3 | 4 | 5 | 6 |
| 1 | 1.20 (77.9%) | 1.80 (14.9%) | 2.58 (2.8%) | 3.10 (0.7%) | 2.33 (0.2%) | 3.00 (0.1%) |
| 2 | 1.22 (2.6%) | 2.14 (0.5%) | 1.00 (0.1%) | 2.00 (0.1%) | - | - |
| 3 | 1.25 (0.3%) | - | - | - | - | - |

*Table 3: Average number of SDSs in relation to numbers of companies and universities in a co-author group; percentage of total publications indicated in parentheses*

The analysis reveals that, on average, the number of SDSs present increases with the number of universities involved in the collaboration (from 1 through to 4 universities, when there is one company as co-author). This phenomenon could be because private companies are searching out multiple universities to attain sets of capabilities that a single university is not able to provide (due to lack of competencies or resources; because of inactivity in a particular area of research; because researchers that could offer competencies are already employed on other projects) or because of other reasons that a single university does not conform with the company's expectations and needs.

**5.2 Frequencies of university collaborations by companies**

Given the heterogeneity of the companies involved (in terms of competitive sector, technological area, size, strategy, etc.) various behaviors can be expected in their university collaborations. We observe that 51% of the total 509 firms involved collaborated only once in the period under consideration. This resulted in a total of 261 single-event collaborations, or 13% of the total (Table 4).

| No. of collaborations per single company | No. of companies | % of the No. of companies | % of the cumulative No. of companies | Sub-total of collaborations | % of the total No. of collaborations | % of the cumulative No. of collaborations |
|---|---|---|---|---|---|---|
| 1 | 261 | 51.3 | 51.3 | 261 | 13.2 | 13.2 |
| 2 | 87 | 17.1 | 68.4 | 174 | 8.8 | 21.9 |
| 3 | 41 | 8.1 | 76.4 | 123 | 6.2 | 28.1 |
| 4 | 20 | 3.9 | 80.4 | 80 | 4.0 | 32.2 |
| 5 | 22 | 4.3 | 84.7 | 110 | 5.6 | 37.7 |
| 6 | 12 | 2.4 | 87.0 | 72 | 3.6 | 41.4 |
| 7 | 9 | 1.8 | 88.8 | 63 | 3.2 | 44.5 |
| 8 | 4 | 0.8 | 89.6 | 32 | 1.6 | 46.1 |
| 9 | 5 | 1.0 | 90.6 | 45 | 2.3 | 48.4 |
| 10 | 7 | 1.4 | 91.9 | 70 | 3.5 | 51.9 |
| More than 10 | 41 | 8.1 | 100 | 953 | 48.1 | 100 |

*Table 4: Intensity of university-company collaborations per company*

Most companies had 10 collaborations or less: these companies represent 92% of all those in the census and contributed to roughly 52% of all collaborations.

The analysis of the intensity of collaborations by single companies reveals that 8% of collaborating companies account for 48% of all collaborations. The 9 companies[10] most active in collaboration with universities realize 23% of the overall collaborations. The most active company participated in 83 collaborations and the ninth most active participated in 25 collaborations. Seven of these companies operate in the pharmaceutical sector and two in electronics.

**5.3 Geographic distances: analysis of frequency**

As stated above, 1,534 publications were produced by 1,226 university-company pairs, working in 1,983 collaborations. Figure 1 presents the frequency curves for these collaborations, in terms of geographic distance[11] of the partners, in intervals of 50

---

[10] In this case the company is considered as a whole, not taking account of possible dispersed locations.
[11] Geographic distance is expressed as the straight-line surface distance between partners, measured by means of coordinates of longitude and latitude.

kilometers (km).

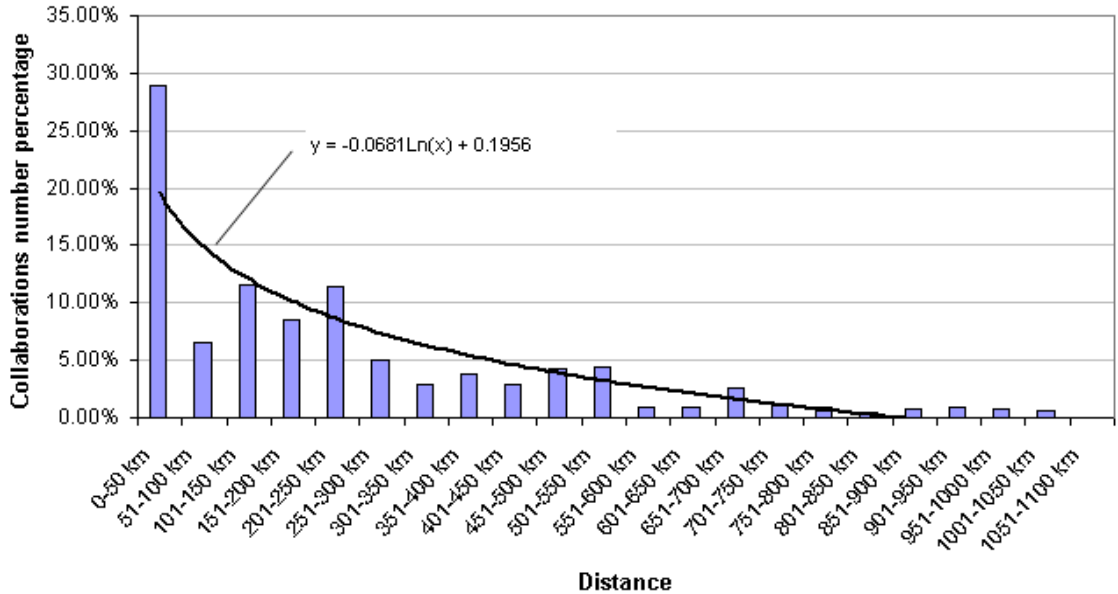

*Figure 1: Collaborations (percentage values) grouped by ranges of 50 km*

Almost 29% of collaborations occur between partners located within a range of 50 km, while 67% of collaborations occur between partners within 250 km of one another.

The average distance for all university-company collaboration pairings is seen to be 255 km. The maximum distance registered is 1,048 km.

## 6. Assessment of proximity effect

The preceding section described the general nature of university-company research collaborations. This section is intended to determine whether a geographical proximity effect occurs in companies' choice of university partners. In the past, the influence of geographic proximity on research collaboration has been studied mostly by means of spatial interaction approaches. In particular, the well-known gravity model (Ponds et al., 2007, Maggioni and Uberti, 2007) is used for predicting and analyzing the interaction between two systems (such as regions). This model assumes that the collaboration has no other cause than the influence of the "masses" of the two interacting systems. The approach adopted by this study is more articulated. We proceed by steps, introducing additional information concerning the SDSs that would be relevant to the selection of a partner but which appears to be increasingly less known to company decision-makers. At each step we calculate the expected distance between partners in light of the additional information and compare this with the real distance.

### 6.1 Calculation of expected average distance and comparison to real distance

For each SDS-company collaboration, the real distance between the two organizations involved is calculated. The average real distance for all SDS-company collaborations is seen to be 226 km[12].

At this point, we consider each SDS of a specific university as a geographical reference point, characterized exclusively by its territorial location. For each collaboration and SDS-company pair, we calculate the average distance between the company and all the universities with scientists belonging to the same SDS.

We begin our analysis by assuming that the company's choice is purely casual. Without additional information regarding the SDSs and considerations of an economic nature, the probability that a company will choose university $i$ is the same as that for all other universities. We therefore measure the expected average distance of the total of all the collaborations. This amounts to 398 km, 76% higher than the real average distance. The analysis at the level of each SDS shows that in 1,868 out of 2,363 SDS-company collaborations (79%), the real distance is less than that expected. This leads us to believe that the company's selection of partners has been influenced by cost-benefit criteria or other information.

**6.2 Calculation of barycentric distance with respect to mass; comparison to real distance**

We now introduce information regarding the scientists belonging to the SDS of interest to each university. In the absence of further information and economic considerations, the capacity of a university to attract a company should increase with the increase in the number of its scientists specialized in the SDS of interest. For each SDS-company collaboration, the university is no longer represented by a geographical point, rather by a circle whose area is proportionate to the number of scientists belonging to the SDS in question. The geographic distance is thus weighted by a size factor. We term it "barycentric distance with respect to mass", with the formula:

$$MBD_{jw} = \frac{\sum_i m_{iw} * d_{ij}}{\sum_i m_{iw}}$$

where:
- $MBD_{jw}$ = barycentric distance with respect to university mass, between the private company $j$ and the Italian university system, for SDS $w$;
- $m_{iw}$ is the "mass" of SDS $w$ of the university $i$, measured as number of researchers in SDS $w$ at university $i$ (for the triennium in consideration);
- $d_{ij}$ is the distance between company $j$ and university $i$.

The summations include all universities with at least one scientist identified in the SDS under consideration.

Following an analysis that is analogous to that of the previous section, the average barycentric distance with respect to mass is seen to be 376.3 km, which is 66% higher than the real average distance. Of the total of 2,363 SDS-company collaborations, 1,833 (77.6%) show a barycentric distance, with respect to university size, thatgreater than the true geographic distance. This denotes the presence of a proximity effect and/or influence of other cost-benefit criteria and information in the decision-making process of the private company.

---

[12] Such distance, which refers to SDS-company pairs is slightly different from that reported in paragraph 4.3, which referred to university-company pairs.

**6.3 Calculation of barycentric distance with respect to Scientific Strength; comparison to real distance**

Finally, we introduce information on the research performance of the SDS of each university. This information, which was not available to the industrial sector during the period under observation, is taken from the ORP. As mentioned in section 4, ORP uses specifically developed algorithms to arrive at a listing of the scientific publications of individual Italian university scientists (about 36,000). The research performance of any given university in each SDS is measured by aggregation of data relating to the individual scientists that belong to that particular SDS. As a performance indicator we use the "Scientific Strength" (SS) of the university in the pertinent SDS[13]. Scientific strength is a compound indicator defined as the number of publications by a university in an SDS, with each publication weighed for the prestige (Impact Factor, IF) of the publishing journal[14]. It follows that: i) the higher the research output in an SDS (i.e. the higher the number scientists in an SDS, if the average productivity across universities remains the same) and/or ii) the higher the prestige of the publishing journals, the higher the value of SS. The consideration of scientific journal publication as the sole output of research, excluding other recognized outputs such as proceedings, monographs, patents or prototypes, receives ample justification in literature. A reaffirmation is that in the first national evaluation of research for Italy (VTR-CIVR, 2006), which took place during the same triennium under examination in this study, journal articles represented a minimum of 85% and a maximum of 99% of all products submitted for evaluation by universities. In the scientific disciplines considered, journal publications represented over 90% of the total products presented.

Expressed as a formula, the SS of university $k$ in sector $s$ is:

$$SS_{ks} = \sum_i IF_i \cdot b_{iks}$$

$$b_{iks} = \begin{cases} 1, \text{ if SDS } s \text{ at university } k \text{ featured at least one author among those for the publication } i \\ 0, \text{ otherwise} \end{cases}$$

$IF_i$ = Percentile rank of the journal's impact factor (in the distribution for its SCI scientific category) publishing publication $i$

Analogous to the previous section, the expected barycentric distance with respect to the Scientific Strength was compared to the real distance.

The results of the analysis confirm the preceding picture. The expected SS barycentric distance is seen to be 357.8 Km, 58% higher than the real average distance.

---

[13] For further information on the calculations and the precautions taken in the use of this indicator see Abramo et al., 2008b.

[14] The authors are aware of the limitations in this use of "impact factor" (Moed and Van Leeuwen 1996; Weingart 2005), but the number of citations to single articles was not available in our data source.

Of a total of 2,363 SDS-company collaborations, a full 1,809 (76.6%) present an expected SS barycentric distance that is greater than the real distance.

A private company, as a rational economic decision-maker in the absence of asymmetric information, should logically choose a university partner that presents the best trade-off between scientific quality and cost (geographic distance). The fact that the introduction of information on the quality of research of single SDSs still produces results that are similar to those seen in the preceding section may signal the presence of information asymmetry in Italy, leading us to conclude that geographic proximity is the preponderant factor in companies' choice of research partners. Table 5 synthesizes the results of the step by step analyses with increasing levels of information. One can see that with the increase in information on the SDSs that are the subject of collaborations, the expected distance decreases and comes increasingly closer to the real distance with every step. But in each case, the average expected distance is always higher than the real figure, which confirms the existence of the geographical proximity effect.

| Distance | Value (km) | Ratio to real average distance | Used information |
| --- | --- | --- | --- |
| Real average distance | 226.04 | 1 | None |
| Expected average distance | 397.92 | 1.76 | Geographic |
| Expected mass barycentric distance | 376.29 | 1.66 | Geographic + SDS mass |
| Expected SS barycentric distance | 357.75 | 1.58 | Geographic + SDS performance |

*Table 5: SDS-company research collaboration distances*

## 7. Assessment of market efficiency

The observations in the previous section indicate the presence of a proximity effect in which geographic distance is the main factor for selecting a university partner, at the expense of both the mass and the scientific strength of the university in the pertinent SDS. This phenomenon may occur for two reasons. The first possible explanation is that companies really lack information on the relative research quality of universities and research groups. The second is that they place more emphasis on the costs associated with distance than they do on quality. If this second explanation holds true we should find no higher quality university or individual scientist/research group closer than the one actually chosen by the company. In this section we use information about research performance of universities and single scientists to assess whether a company could have chosen a higher quality university or higher ranked scientist within a shorter distance.

### 7.1 Analysis of the choice of university

For every SDS, the "Qualitative Productivity" (QP) of each university was calculated. This is an indicator that measures the efficiency of the research activity as a ratio between output (Scientific Strength of a university in a given SDS) and input (number of scientists in the pertinent SDS)[15]. As a formula, the Qualitative Productivity

---
[15] For further information on the calculations and precautions taken in the use of this indicator see

of university *k* in SDS *s*:

$$QP_{ks} = \frac{SS_{ks}}{Add_{ks}}$$

$SS_{ks}$ = Scientific Strength of university *k* in SDS *s*

$Add_{ks}$ = number of university research staff associated with SDS *s* at university *k*

For each SDS, the QP indicator was used to develop a ranking of all Italian universities according to research quality. In the analysis of the efficiency of the company's choice among these potential partners, we consider only co-authored publications involving a single company[16]. These represent 1,480 (96.5%) out of the total of 1,534 co-authored publications. In 1,382 of these cases, or 93.8% of the total, the private company could have chosen an Italian university with a superior ranking of Qualitative Productivity[17]. On average, one can see that for each SDS-company collaboration, there would be 11 universities better than the one chosen, that is to say 28.4% of the universities active in the SDS under examination. Thus in the majority of cases, private companies are not selecting research collaborations with the universities most qualified in the pertinent sector. We would expect that the universities that are better than those chosen would be found at a greater distance from the company, which would justify the choice from a cost-benefit point of view. However calculations show that in 58.5% of the cases (809 out of 1,382 research publications) there is at least one university, geographically closer to the company that has a better performance in the pertinent SDS compared to the university actually chosen. This proves that companies lack complete information on the relative research quality of their research partners. The high percentage of choices in favor of lesser quality universities might be explained, in part, as an undesirable outcome of private companies resorting to social proximity as a substitute for information on quality. In 54.66% of cases (809 out of 1,480) the choice results as being inefficient both for the quality and the cost (distance) dimensions. This high percentage gives an indication of the extent to which information asymmetry creates inefficiency in the market for public-private research collaboration.

**7.2 Analysis of choice of individual researchers**

It seems likely that a company planning collaboration could make its selection by searching for a research group or a single scientist, rather than for a university. In order to examine this possibility, an analysis similar to the one above was conducted at the level of single university researcher. For this analysis we used the indicator of Scientific Strength, defined in section 6.3, to rank university scientists in each SDS.

Of the total of 1,400 publications considered in the analysis[18], there are 1,333 cases

---

Abramo et al., 2008b.

[16] In the presence of more than one company, it is not possible to identify the "decision-maker".

[17] Where more than one university is involved in a university-industry collaboration grouping, the study conservatively considers the university with the highest position in the ranking for the SDS at stake.

[18] For the analysis of the companies' choice of researchers, the study considers publications co-authored by a single private company (excluding more complex dynamics) and one or more universities. To further increase the reliability of the procedure, the analysis considers only the publications (1,400) by the 2,562

(95% of the total) in which the private company could have collaborated with another researcher in the same SDS who was better classified than the one actually chosen. Again, considering the variable of distance, there are 877 cases out of 1,333 (65.8%) in which there was at least one geographically closer researcher with superior scientific performance, in the same SDS, compared to the researcher actually chosen.

The results of the analysis at the single scientist level are less robust than the preceding analysis, since the analysis at the level of the single researcher requires two assumptions. The first is that potential university partners would not refuse a proposal for collaboration advanced by a company. This is a reasonable assumption to make in the Italian context, where universities are starving for financial resources[19]. The second assumption is that all university scientists in the same sector of specialization (SDS) are perfect substitutes for one another, an assumption that is as realistic as the sector is narrow[20].

Given that in each SDS there are on average 73 scientists with a performance that is superior to that of the best of the co-authors, and 22 with both a superior performance and a closer geographic distance, the probability that all refused or none possessed the competences requested by the company is rather low. We thus feel quite confident in stating, as previously illustrated at the level of choice of universities, that analysis at the level single scientists confirms the presence of information asymmetry as a cause for inefficiency in the market for public-private research collaboration.

## 8. Conclusions

National markets for public-private research collaborations are likely to be characterized by information asymmetry. Italy is an example of a nation with an entirely public university system, where national evaluation of research has only recently been introduced. One could expect to find asymmetry problems, and indeed the phenomenon seems quite noticeable. This study attempts to assess whether there is inefficiency in the market for public-private research collaboration, stemming from information asymmetry..

The results demonstrate that in the majority of cases the collaborations developed between universities and companies are "exclusive" in character: approximately 78% of collaborations feature the participation of only two partners. It is very rare (3.5%) that more than one company is involved in a collaboration group. In contrast, joint research by more than one university with a single private company is more frequent (present in 19% of cases). This observation is consistent with the missions and typologies of the different actors involved: universities are dedicated to maximizing diffusion of knowledge, while private companies attempt to use knowledge for their own specific commercial purposes.

Furthermore, one can see that with the increase in the number of universities involved in collaborations, there is also an increase in the number of disciplinary sectors

---

scientists that remained affiliated to the same university and the same SDS throughout the triennium under consideration, out of a total of a potential 2,966 university researchers.

[19] On average, 90% of government funding (which represents 90% of the entire university budget) is required to cover employees' salaries.

[20] The 183 SDSs that we consider in our analysis is more than the number of Science Citation Index categories (168) in the same disciplines.

of university researchers. The involvement of more than one university probably stems from companies' needs to access different competencies that are not available, at least in the manner desired, in a single university.

The study examined the geographic distance involved in collaborations. Approximately 29% of university-industry collaborations developed between partners located within a 50 km range, while 67% occurred with partners that were within 250 km.

A comparison was made between real company-to-university distances and the expected barycentric distance for both university size (mass) and Scientific Strength. This showed the presence of a proximity effect and the importance of the geographic distance factor in the choice of university partner.

It was further noted that, in the majority of cases, private companies do not develop research collaborations with universities that are the most qualified in the pertinent discipline. Upon examination of the classification of universities by "Qualitative Productivity" in each scientific disciplinary sector, we see that in 93% of cases companies could have collaborated with a higher ranking Italian university.

Considering cost-benefit analysis, the universities that were better qualified than those actually chosen for collaboration should be found at a greater distance from the private companies. Instead, findings show that in 54% of cases there was at least one university closer to the company that had a higher ranking in the pertinent SDS, when compared to the university that was actually chosen.

The same type of analysis, conducted at the level of single university scientists reveals that in 95% of cases the private company could have collaborated with another scientist in the same SDS who was better classified than the one actually chosen. An examination of this data, together with the distance variable, shows that in 65% of cases there was at least one researcher, with superior scientific performance, affiliated to a university that was closer to the company.

The inefficiency in choice of university partner in terms of quality and cost (distance) indicates that there is truly inefficiency in the market for public-private research collaboration. It is possible that the less than optimal choice of company is due to the fact that nearest best scientist refused the request for collaboration (an improbable hypothesis given the scarcity of research funds for Italian universities), or else the specialization of the scientist did not correspond exactly to that required by the company. However the sheer number of top scientists nearer the companies leads one to think that the occurrence of the above hypotheses, however possible they may be, is in fact improbable. It is also true that information on the quality of research by public scientists is not available in Italy, which obviously does not favour an efficient choice. Distance certainly assumes a fundamental role in developing collaborations but certain other factors must also play an important role, apparently more so than the factor of excellence of university partners. It is likely that some of these factors cannot be observed exclusively through bibliometric analysis. Bibliometric analysis does not capture all the aspects of such a complex phenomenon as research collaboration, which can stem from ties of personal familiarity, or from the necessity of involving specific partners to access specific financing, etc. One can certainly assert, however, that bibliometric types of information could assist companies to better their decision-making processes. It is exactly the quantitative-qualitative nature of bibliometric data that could assist in reducing the information asymmetry between the industrial world (demand) and the university sphere (supply). The use of bibliometric information by private companies would likely lead to more effective and productive collaborations, with the potential to guarantee greater benefits to the parties involved, and to the community as a whole.